\documentclass[a4paper,10pt,twocolumn]{article}
\usepackage{graphicx}
\usepackage{subfigure}
\usepackage{multirow}
\usepackage[english]{babel}
\usepackage{ucs}
\usepackage[latin1]{inputenc}
\usepackage{multicol}
\usepackage{float}
\usepackage{cite}
\usepackage{url}
\usepackage{amsmath}%
\usepackage{amsfonts}%
\usepackage{amssymb}%

\begin{document}

\twocolumn[
 \begin{center}
 {\Large\textbf{Geant4 Simulation of a filtered X-ray Source for Radiation Damage Studies}}\\ \medskip
\textsc{M.~Guthoff, O.~Brovchenko, W.~de~Boer, A.~Dierlamm,  T.~M\"uller, A.~Ritter\footnotemark, M.~Schmanau, H.-J.~Simonis\\
{\it  Institut f\"ur Experimentelle Kernphysik, Karlsruhe Institute of Technology, Campus S\"ud,
P.O. Box 6980, 76128 Karlsruhe, Germany\\
}}
\bigskip

{\bf\large Abstract}\\
\end{center}

Geant4 low energy extensions have been used to simulate the X-ray spectra of industrial X-ray tubes with filters for removing the uncertain low energy part of the spectrum in a controlled way. 
The results are compared with precisely measured X-ray spectra using a silicon drift detector.
 Furthermore,  this paper shows how the  different dose rates in  silicon and silicon dioxide layers of an electronic device can be deduced from the simulations.
\bigskip 
]
\footnotetext{Now at Max-Planck-Institut f\"ur Physik, Halbleiterlabor, Otto-Hahn-Ring 6, D-81739 M\"unchen, Germany.}

\section{Introduction} 
Radiation hardness of semiconductors can be tested by irradiation with X-rays, since these introduce efficiently trapped ionization in the silicon oxide layers, which e.g. leads to voltage shifts in MOSFET transistors.  The damage depends not only on the X-ray flux, but also strongly on the spectrum of the X-rays. The total flux can be obtained from a calibrated pin diode, but an accurate measurement of the X-ray spectrum is complex and time consuming. A simulation is a more convenient way to get the spectrum. In this paper Geant4\cite{Geant4} low energy extensions have been tested for this purpose. 
The results are compared with precisely measured X-ray spectra using a silicon drift detector (SDD).
 Furthermore,  the different dose rates in  silicon and silicon dioxide layers of an electronic device can be deduced from the simulations.
 
\section{Experimental Setup}

\begin{figure}
	\centering
		\includegraphics[width=0.45\textwidth]{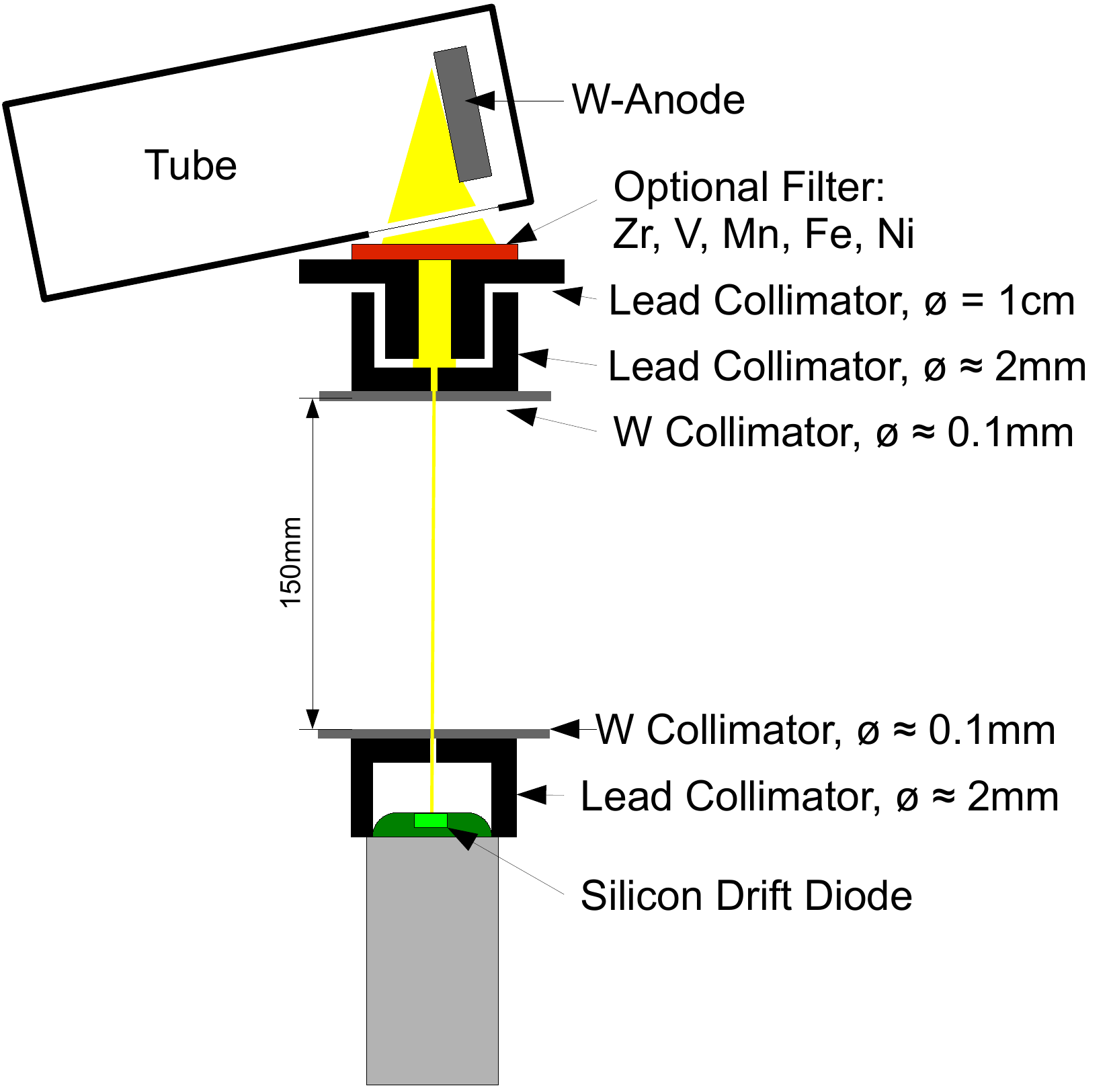}
	\caption[Experimental Setup]{Sketch of the experimental setup. The lead collimators are used to mount the tungsten collimators and reduce background radiation.}
	\label{fig:exp_setup}
\end{figure}

The X-ray tube of type DX-W8x0.4-2S with a tungsten anode has a maximum power of 2000\,W and a maximum voltage of 60\,kV. The tube window is made of 0.4\,mm Beryllium, so that only low energy photons are absorbed by the window. The X-ray tube is equipped with different filters that can be mounted behind the exit window. These filters are vanadium (15\,$\mu$m), manganese (25\,$\mu$m), iron (15\,$\mu$m), nickel (15\,$\mu$m) and zirconium (75\,$\mu$m). These filters absorb  low energy photons in different energy regions.
The minimal configurable current of the HV-module is 2\,mA. The top part of Fig.~\ref{fig:exp_setup} shows the tube with the filters. The tungsten collimator for measuring the spectrum at a reduced rate in a silicon drift detector is shown in the lower part.

\section{Dosimetry}
\label{sec:dosimetry}
One critical information in the test for radiation hardness of semiconductor devices is the dose. This chapter describes the way of calculating the dose to any kind of material if the X-ray spectrum and a reference measurement are known. 
\subsection{Dose rate calculated from X-ray spectrum.}
The X-ray energy spectrum provided by the source is given as $N(E)$ which can be either measured or simulated. Since obtaining a flux normalized spectrum is even more difficult, $N(E)$ is arbitrary scaled. Only a fraction of the spectrum is absorbed during irradiation. The absorbed part is determined by:
\begin{equation}
N_{abs}(E) = N_{Filter}(E) \cdot (1 - e^{\rho  \cdot  \sigma(E)  \cdot  d}),
\label{eq1}
\end{equation}
where $\rho$ represents the density, $\sigma$ the absorption coefficient and $d$ the thickness of the material.
The absorbed power can be calculated to:
\begin{equation}
		P_{abs}	= \alpha \cdot \int_0^{E_{max}} N_{abs}(E)  \cdot  E  \cdot  dE,
		\label{equ:Pabsspec}
\end{equation}
where $\alpha$ represents the unknown scaling factor.
The dose rate is defined as:
\begin{equation}
	\dot{D} = \frac{P_{abs}}{m}.
	\label{equ:dose}
\end{equation}

\subsection{Absorbed dose in a depleted diode as reference}
The missing parameter $\alpha$ can be obtained with a reference measurement. With a depleted silicon diode it is possible to measure the photo current from ionizing radiation. Each generated electron-hole pair represents $3.65\,eV$ \cite{SI-IoE}, so that the absorbed power in the device can be calculated as:
\begin{equation}
	P_{abs_{Si}} = \frac{I_{photo}}{e}*3.65 eV,
	\label{equ:Pabsmeas}
\end{equation}
where $e$ denotes the elementary charge.\\
The dose rate in this device is then given by
\begin{equation}
	\dot{D}_{Si} = \frac{P_{abs_{Si}}}{m_{Si}},
\end{equation}
where $m_{Si}$ is the mass of the diode. The measured dose rate is only valid for silicon material of the same thickness. In our measurements we used a planar diode with a thickness of $300\,\mu m$ and a lead collimator with a circular opening of $\varnothing = 1.5\,mm$. Depletion over the whole thickness was ensured via a CV-measurement.

Using Eqs.~\ref{equ:Pabsspec} and~\ref{equ:Pabsmeas} one can now obtain $\alpha$:
\begin{equation}
	\alpha = \frac{I_{photo} \cdot 3.65 eV}{e \cdot  \int_0^{E_{max}} N_{abs_{Si}}(E)  \cdot  E  \cdot  dE},
	\label{equ:alpha}
\end{equation}
where $N_{abs_{Si}}$ is the absorbed spectrum in the reference diode.
It is now possible to calculate the dose for any other material if the energy spectrum of the X-ray source is known. The spectrum can either be measured or simulated. Both ways and a comparison are described in Sect.~\ref{sec:spectrum}.
\begin{table*}
	\centering
		\begin{tabular}{|c|c||c|c||c|c||c|c||c|}
		\hline
\multicolumn{2}{|c||}{\textsuperscript{29}Cu}&\multicolumn{2}{|c||}{\textsuperscript{47}Ag}&\multicolumn{2}{|c||}{\textsuperscript{56}Ba}&\multicolumn{2}{|c||}{\textsuperscript{65}Tb}&\textsuperscript{241}Am\\
\hline
\hline
I$_{rel}$&E[keV]&I$_{rel}$&E[keV]&I$_{rel}$&E[keV]&I$_{rel}$&E[keV]&E[keV]\\
\hline
\hline
100&8.048		&100&22.162		&100&32.193		&100&44.481		&59.5\\
52&8.028			&53&21.990		&54&31.817		&56&43.744		&13.9\\
17	&8.905		&16&24.942		&18&36.378		&20&50.382		&\\
&						&4&25.456			&6&37.257			&7&51.698			&\\
&						&9&24.911			&10&36.304		&10&50.229		&\\
\hline		
\end{tabular}
\caption{List of materials used for the energy calibration an their characteristic X-ray energies. I$_{rel}$ is the relative intensity of the peaks for the given material.}
\label{tab:CalibrationEnergies}
\end{table*}

\subsection{Calculation of the dose rate in a different material}

Silicon dioxide is often used as an insulator in semiconductor devices, which suffers 
 from charge build-up caused by ionizing radiation. Hence, for radiation damage studies it is important to know not only the total dose rate, but also the dose rate in different layers, which usually have different absorption coefficients. We use $SiO_2$ as an example to show the calculation of the dose rate in any material.

The absorbed power is calculated as in Eq.~\ref{equ:Pabsspec}, and using Eq.~\ref{equ:alpha} one obtains:
\begin{equation}
	\begin{split}
	P_{abs_{SiO_2}}	& = \alpha \cdot \int_0^{E_{max}} N_{abs_{SiO_2}}(E)  \cdot  E  \cdot  dE\\
					& = \frac{I_{photo} \cdot 3.65 eV}{e}\\
					& \quad \cdot \frac{ \int_0^{E_{max}} N_{abs_{SiO_2}}(E)  \cdot  E  \cdot  dE}{ \int_0^{E_{max}} N_{abs_{Si}}(E)  \cdot  E  \cdot  dE}\\
					& =  P_{abs_{Si}} \cdot \frac{\int_0^{E_{max}} N_{abs_{SiO_2}}(E)  \cdot  E  \cdot  dE}{\int_0^{E_{max}} N_{abs_{Si}}(E)  \cdot  E  \cdot  dE} 
	\end{split}
	\label{equ:PSiO2}
\end{equation}
The ratio of the absorbed dose in the reference material and in the target material is therefore used to scale the reference measurement $P_{abs_{Si}}$ to the absorbed power in the target material $P_{abs_{SiO_2}}$.
The dose rate in $SiO_2$ follows from Eqs.~\ref{equ:dose}~and~\ref{equ:PSiO2}:
\begin{equation}
	\begin{split}
	\dot{D}_{SiO_2}	&= \frac{P_{abs_{SiO_2}}}{m_{SiO_2}}\\
					&= \frac{P_{abs_{Si}}}{m_{SiO_2}} \cdot \frac{\int_0^{E_{max}} N_{abs_{SiO_2}}(E)  \cdot  E  \cdot  dE}{\int_0^{E_{max}} N_{abs_{Si}}(E)  \cdot  E  \cdot  dE}.
	\end{split}
\end{equation}
The number of absorbed photons,  $N_{abs_{Si}}$ and   $N_{abs_{SiO_2}}$, can be calculated from Eq. \ref{eq1} using the known energy dependent absorption coefficients $\sigma$ for each material\cite{XCOM}. In principle the method can be used for arbitrary materials and has been applied  for all six possible filter settings.

\section{X-ray spectrum}
\label{sec:spectrum}
A measurement of the X-ray spectrum is often not possible, since the count rates are usually too high for detectors with a suitable energy resolution. Alternatively, the spectrum can be obtained from a simulation. In this chapter we describe both ways, simulations are done with Geant4, the measurements are done with a silicon drift detector.
\begin{figure}
	\centering
		\includegraphics[width=0.45\textwidth]{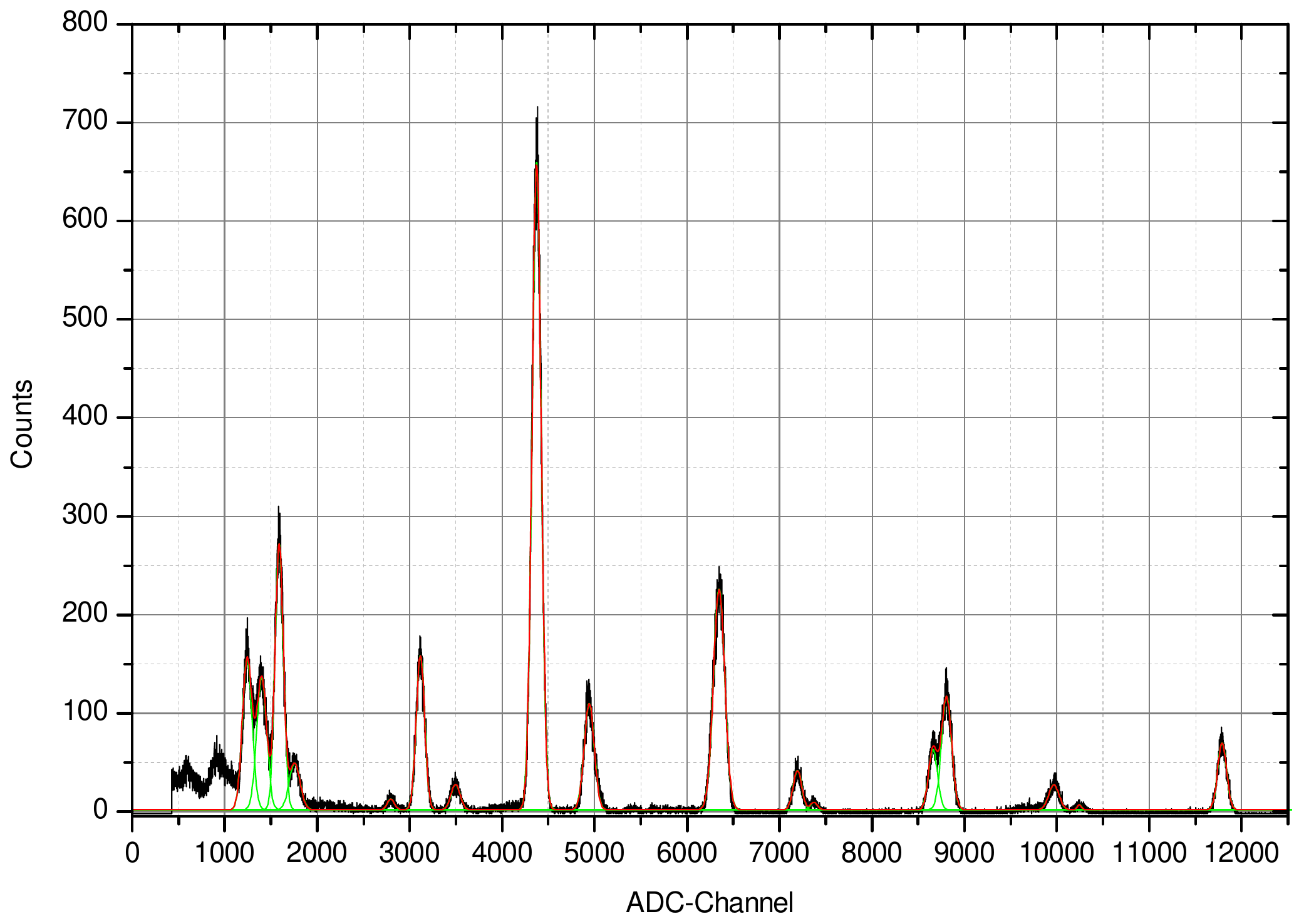}
\caption{Energy deposits from the lines of the materials in Table \ref{tab:CalibrationEnergies}.}
	\label{fig:Ecal1}
\end{figure}
\begin{figure}
	\centering
		\includegraphics[width=0.41\textwidth]{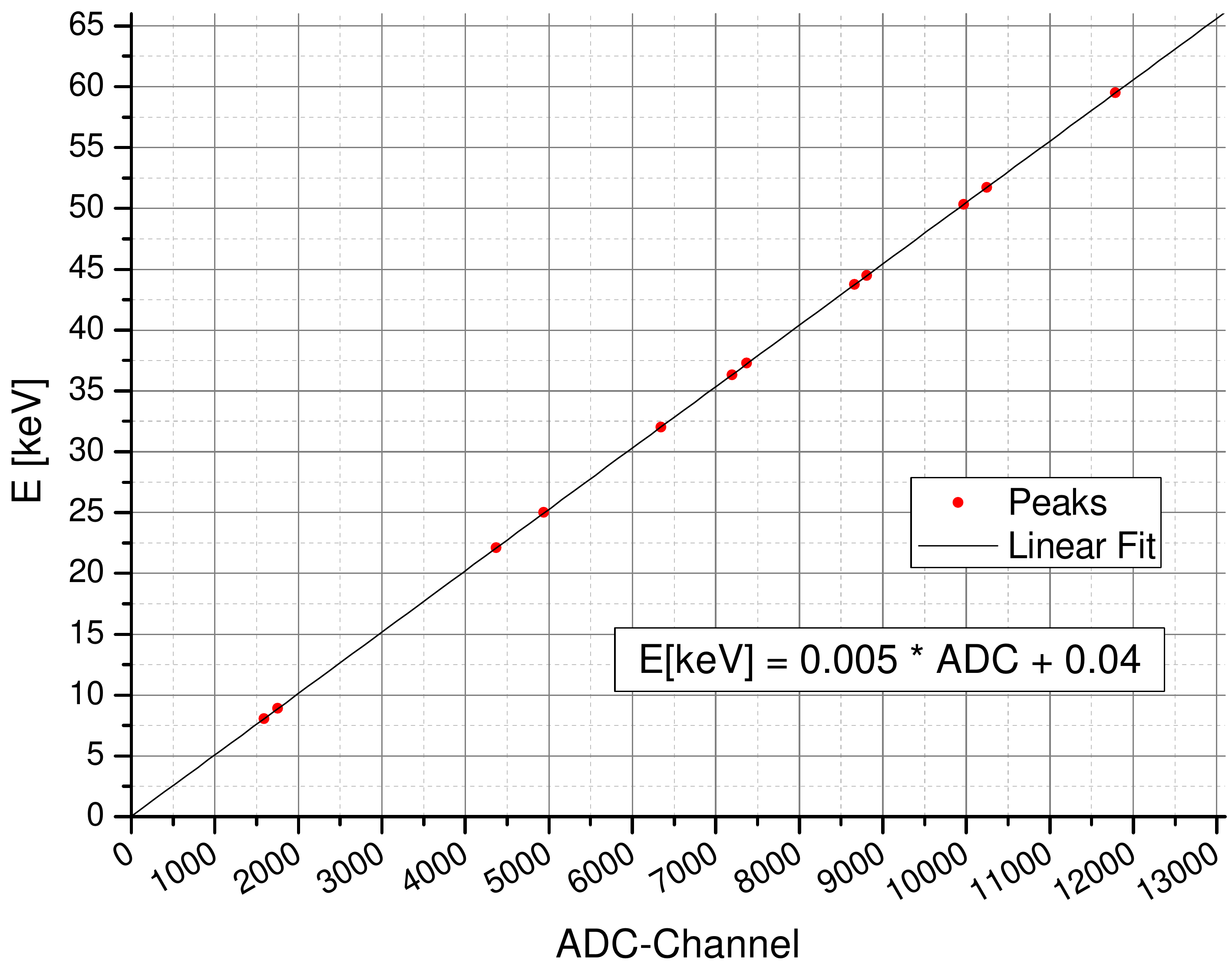}	
	\caption{Linear fit to the known energies of the observed lines in Fig. \ref{fig:Ecal1}.}
	\label{fig:Ecal2}
\end{figure}

\subsection{Measurement of the spectrum}
A silicon drift detector (SDD) with an integrated preamplifier of the type AXAS-P (KETEK GmbH, Munich, Germany) was used as detector. The  signals were further amplified by a shaping amplifier and registered by a readout system Multiport~II (Canberra Industries, Inc., Meriden, CT, USA) with 16384 ADC channels.
All measurements were done with an X-ray energy of 60\,kV and 2\,mA anode current. With this configuration the photon flux in the detector was too high  to detect single photon events, since the flux was several order of magnitude above the  count rate for the slow drift detector. Two tungsten collimators were used to reduce the count rate. These collimators were 1\,mm thick with a hole of $\sim$0.1\,mm diameter. 
 
 One collimator was mounted at the exit window of the tube and another one on the detector, as shown in Fig.~\ref{fig:exp_setup}.

A radioactive \textsuperscript{241}Am source was used for the energy calibration. This source includes different additional materials, which are excited by the \textsuperscript{241}Am source and emit characteristic X-rays. The materials and their X-ray energies can be found in Table~\ref{tab:CalibrationEnergies}.
The relation for the energy calibration is:
\begin{eqnarray*}
E[eV] = (5.04 \pm 0.003) \cdot \textrm{ADC} + (40 \pm 19),
\end{eqnarray*}
as obtained from a linear fit to the observed energy peaks in ADC channels, shown in Fig.~\ref{fig:Ecal2}. The small errors show that the calibration is precise.

The SDD has a thickness of 300$\mu$m. Some photons pass through the detector, especially at higher energies. To compensate for this effect the absorbed spectrum was corrected by weighting it with the inverse photon absorption probability~\cite{XCOM} for a layer of 300\,$\mu$m silicon. 
The energy resolution of the measurement was determined by fitting a Gaussian curve to the L$_{\alpha}$ peak. The full width at half maximum (FWHM) is 0.5\,keV, which is comparable to the resolution 0.58\,keV obtained during the energy calibration.
\begin{figure}
\centering\includegraphics[width=0.45\textwidth]{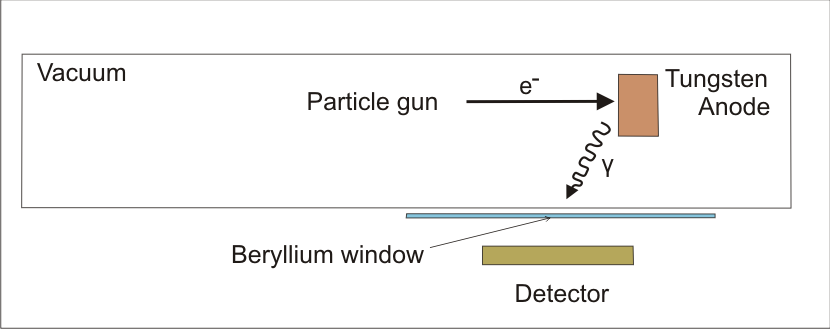}
\caption[Simulation geometry 1]{Geometry for the first simulation step, where all main parts of the X-ray tube are implemented. An electron beam hits a tungsten anode and the resulting photon spectrum is registered by a sensitive air layer (detector).}
\label{Simulation_1}
\end{figure}
\begin{figure}
\centering
\includegraphics[width=0.45\textwidth]{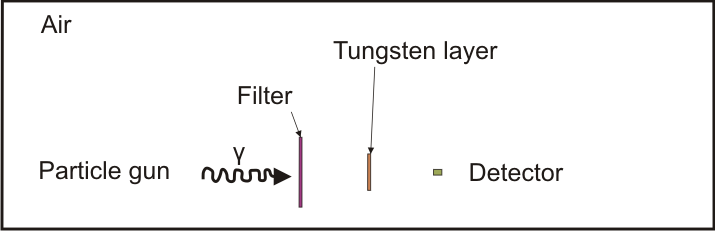}
\caption[Simulation geometry 2]{Geometry for the second simulation step. Photons were generated accordingly to the spectrum  generated by the tube configuration  in Fig.~\ref {Simulation_1}. These photons are registered in a detector after passing a filter and a thin tungsten foil representing the absorption by scattering on the inside of the  collimator in Fig. \ref{fig:exp_setup}.}
\label{Simulation_2}
\end{figure}
\begin{figure}
\centering
\includegraphics[width=0.48\textwidth]{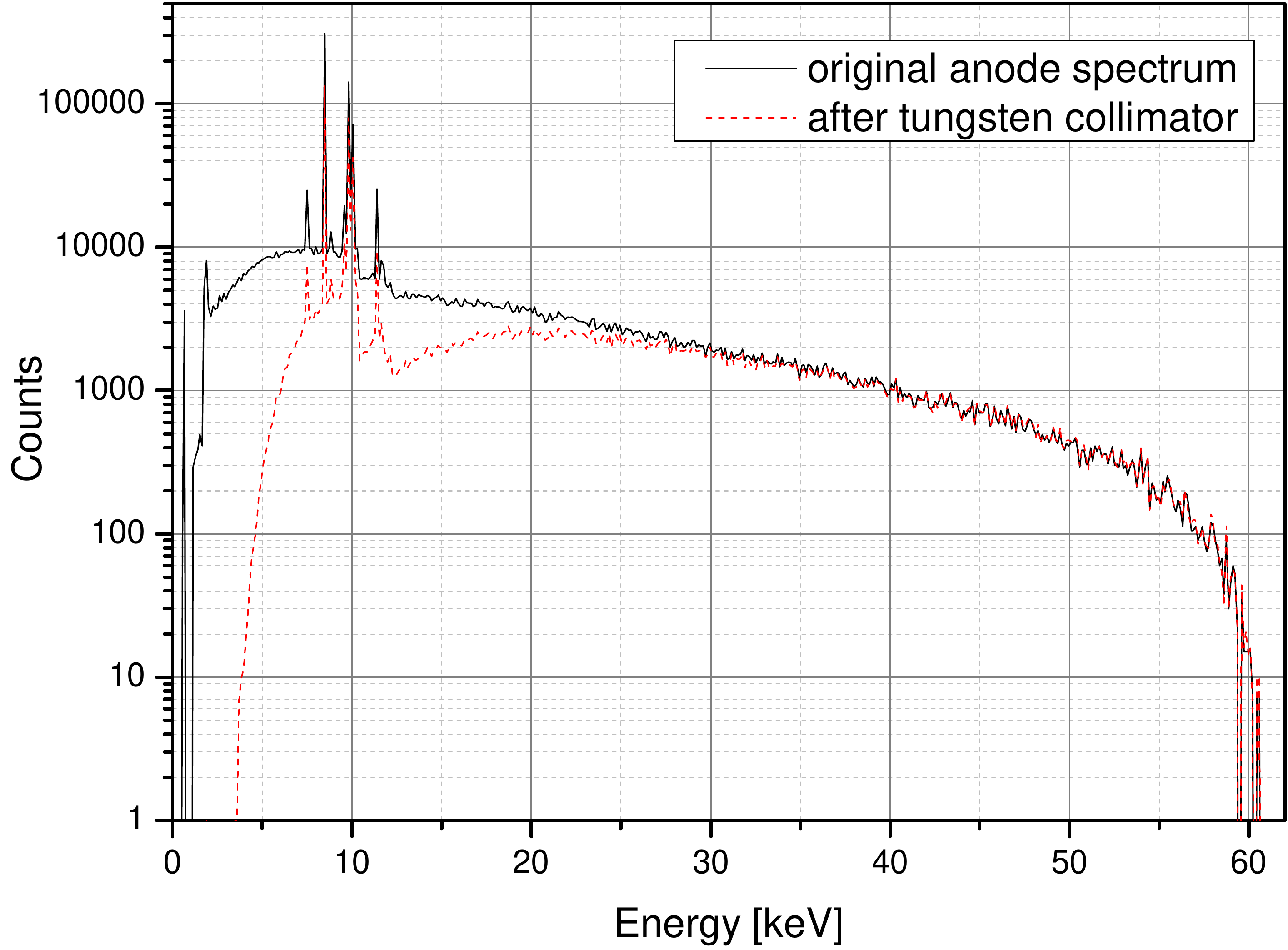}
\caption[Simulated tungsten spectrum]{The simulated spectrum of the tungsten anode using the setup schematically presented in Fig.~\ref{Simulation_1} (upper curve) and the reduction by the thin tungsten foil representing the absorption by the collimator in Fig.~\ref{Simulation_2} without filter (lower curve). This lower curve is compared with data in Fig. \ref{MeasSpectrum_1}. The L$\alpha$, L$\beta$ and L$\gamma$ lines (8.34\,keV, 8.4\,keV, 9.67\,keV, 9.96\,keV, 11.3\,keV) of tungsten as well as the continuum of bremsstrahlung are clearly visible.}
\label{SimSpectrum_1}
\end{figure}

\subsection{Geant4 simulation}
Geant4 is a toolkit developed at CERN for the simulation of the passage of particles through matter~\cite{Geant4}. It is an object-oriented simulation framework which provides a diverse set of software components~\cite{J.Allison2003}. All aspects of the simulation process like the geometry of the system, tracking of particles through materials and external electromagnetic fields as well as the response of sensitive detector components are included in the simulation. Geant4 provides also a set of different physics models to describe the interactions of particles with matter across a wide energy range.
The standard electromagnetic physics package provides implementations of electron, positron, photon and charged hadron interactions. It is suitable for most of the Geant4 applications, but it is not optimized for low energy particles.
Two specific low energy electromagnetic models are available: \textit{Livermore} (based on the Livermore library~\cite{livermore, S.T.Perkins,S.T.Perkinsa}) and \textit{Penelope}~\cite {Penelope}. Geant4 version 9.3.p01 was used and the version of CLHEP libraries was 2.0.4.5.
At first both low energy physics packages were used. However, results obtained with \textit{Livermore} showed differences in comparison with measured spectra. The characteristic tungsten lines had a lower intensity compared to the bremsstrahlung slope. The results presented here are obtained with the \textit{Penelope} package. 

In order to  reduce the simulation time the program was split into two different parts. In the first part the pure tungsten spectrum of the X-ray tube was generated. Then this spectrum was used as input for the second part, where the effect of different filters and collimators could be studied. 
The first step of the simulation is the generation of an electron beam with an energy of $60\pm 1$keV. The  geometry  is shown in Fig.~\ref{Simulation_1}. The environment is vacuum. The vertical position of the beam was varied by 12 mm to simulate the glow filament (the electron source within the X-ray tube). The electrons were shot at a tungsten anode at a distance of 8.5 mm. Below the anode a 0.4 mm thick beryllium window was placed. After the photons, generated at the anode, passed through the beryllium window, their energy was registered by a sensitive air layer (detector).

In the following step this photon spectrum was used as input for the setup shown in Fig.~\ref{Simulation_2}. Material and thickness of the filter were taken from the X-ray tube specifications. After the filter a tungsten collimator was used to collimate the beam. The hole geometry of this collimator is irregular, since  the small hole of $\sim$0.1\,mm was produced by a laser shot. The scattering on the inner edges of the two tungsten collimators was simulated by a thin tungsten foil of 3 $\mu$m placed between the filter and the detector. The thickness was chosen to agree with data, as will be shown in the next section. In Fig.~\ref{SimSpectrum_1} the  simulated  spectra before and  after this thin tungsten layer  are shown.

\begin{figure*}
\subfigure[Measured and simulated spectrum without filter]{
\includegraphics[width=0.48\textwidth]{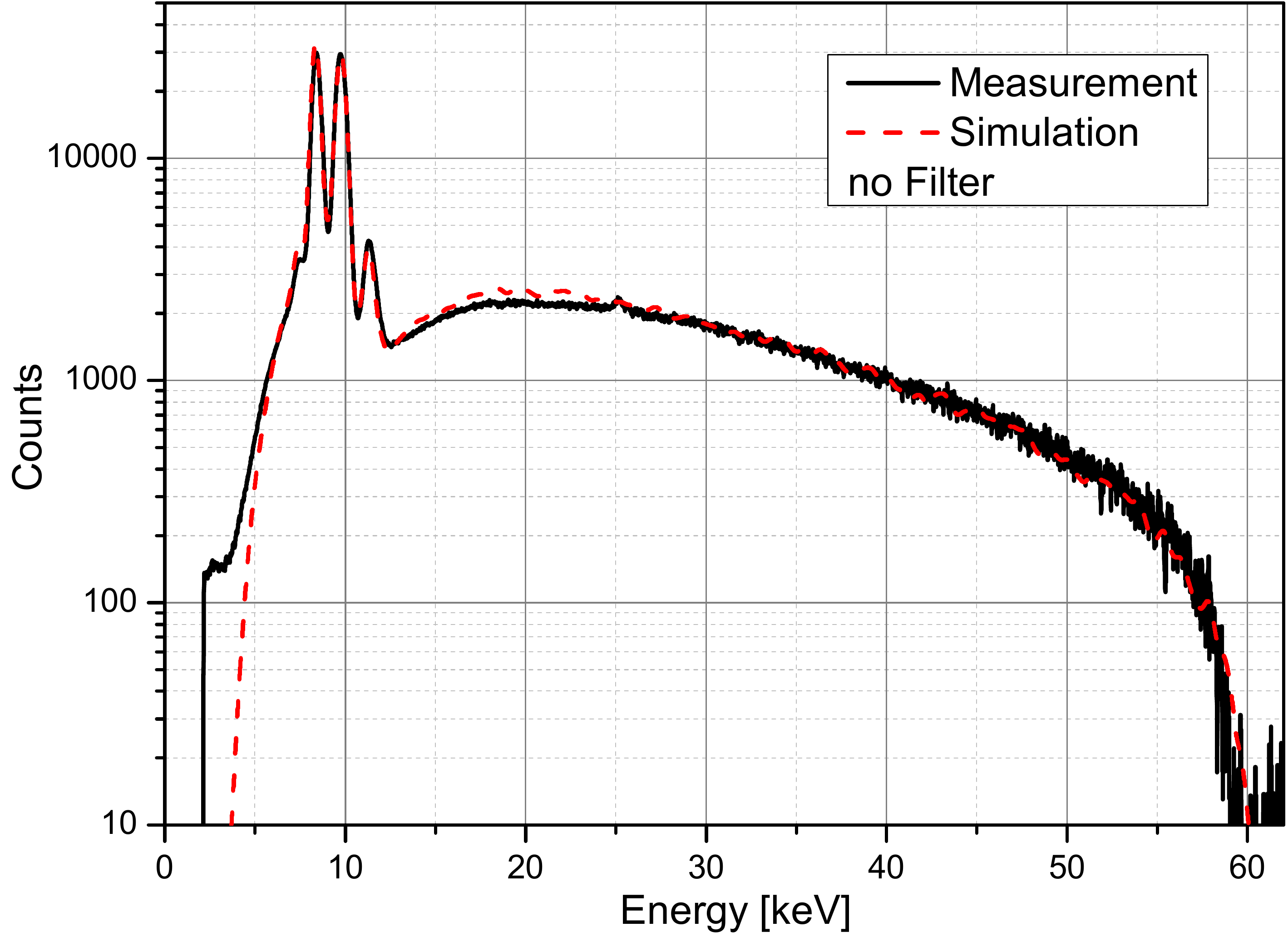}
\label{MeasSpectrum_1}
}
\subfigure[Measured and simulated spectrum after a 75 $\mu$m thick Zr filter]{
\includegraphics[width=0.48\textwidth]{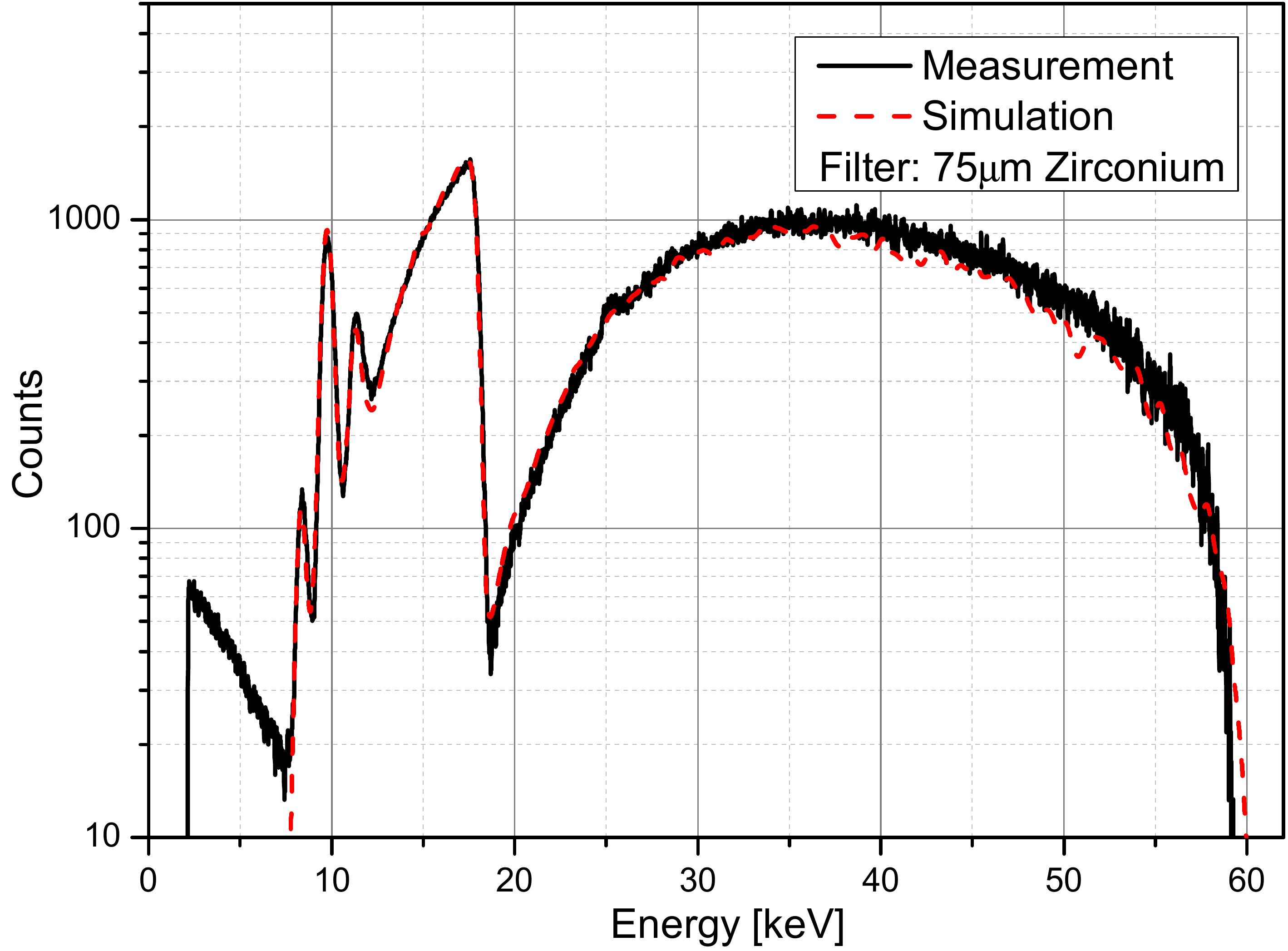}
\label{MeasSpectrum_2}
}
\subfigure[Measured and simulated spectrum after a 25 $\mu$m thick Mn filter]{
\includegraphics[width=0.48\textwidth]{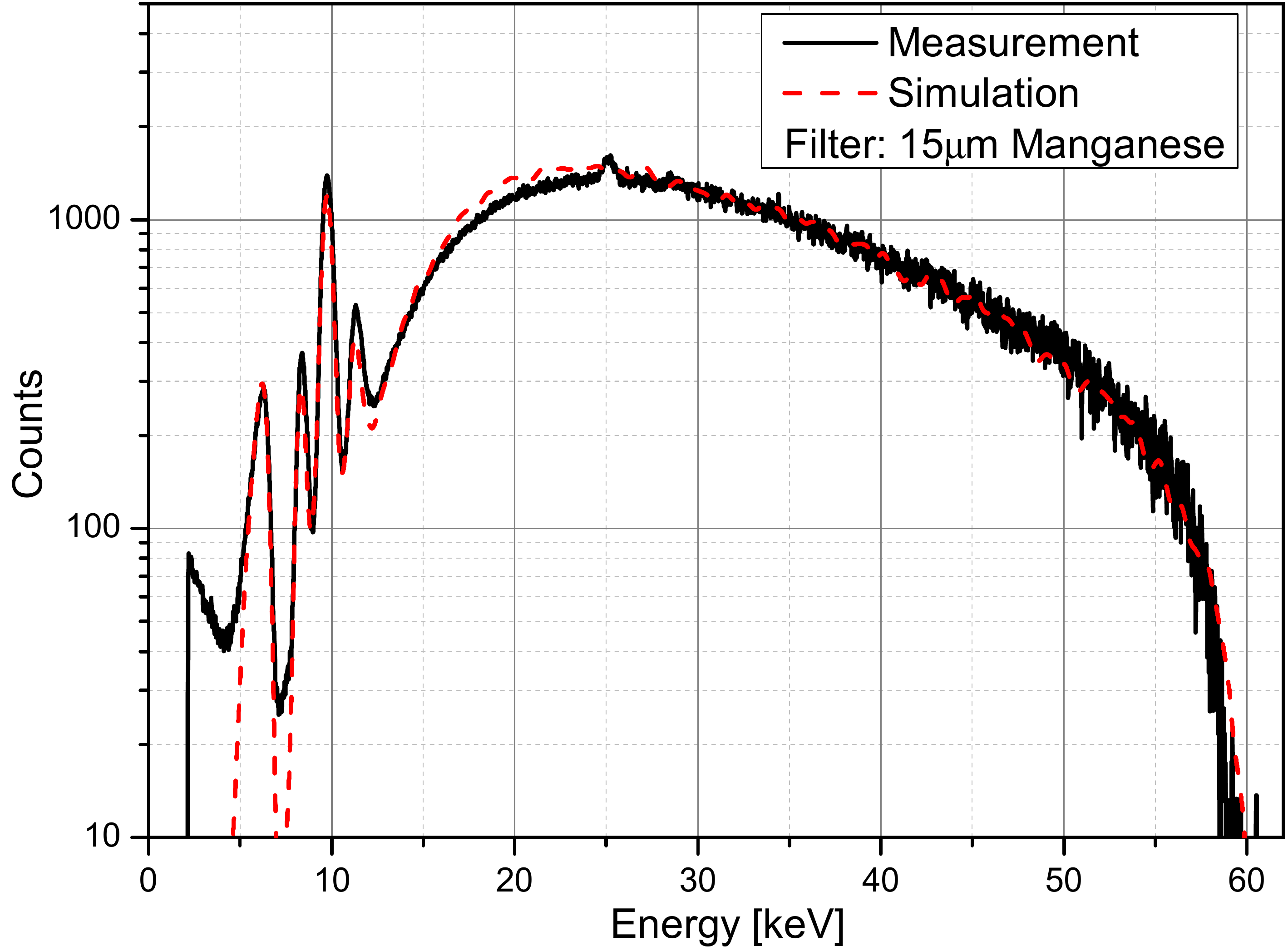}
\label{MeasSpectrum_3}
}
\subfigure[Measured and simulated spectrum after a 15 $\mu$m thick Ni filter]{
\includegraphics[width=0.48\textwidth]{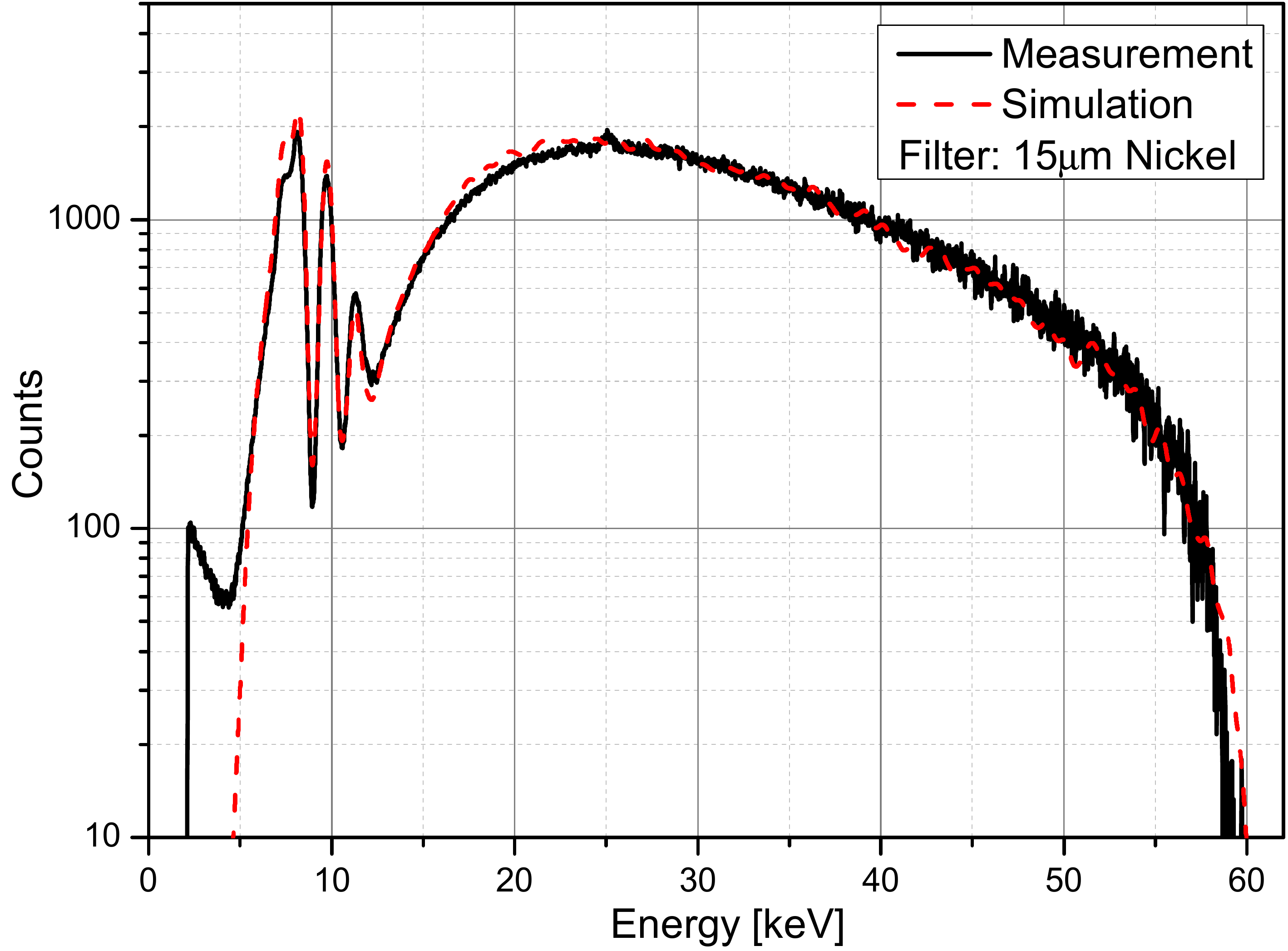}
\label{MeasSpectrum_4}
}
\subfigure[Measured and simulated spectrum after a 15 $\mu$m thick V filter]{
\includegraphics[width=0.48\textwidth]{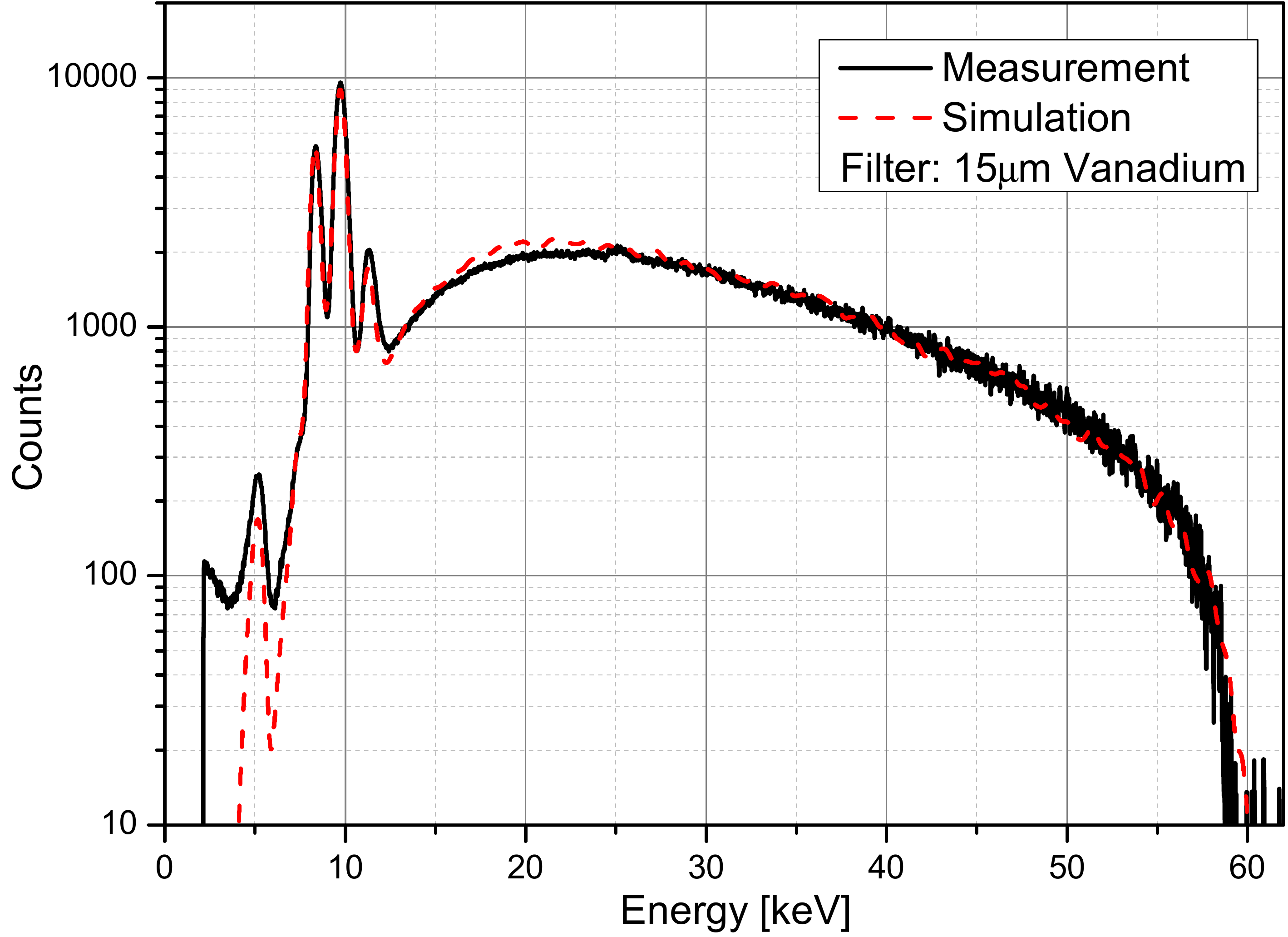}
\label{MeasSpectrum_5}
}
\subfigure[Measured and simulated spectrum after a 15 $\mu$m thick Fe filter]{
\includegraphics[width=0.48\textwidth]{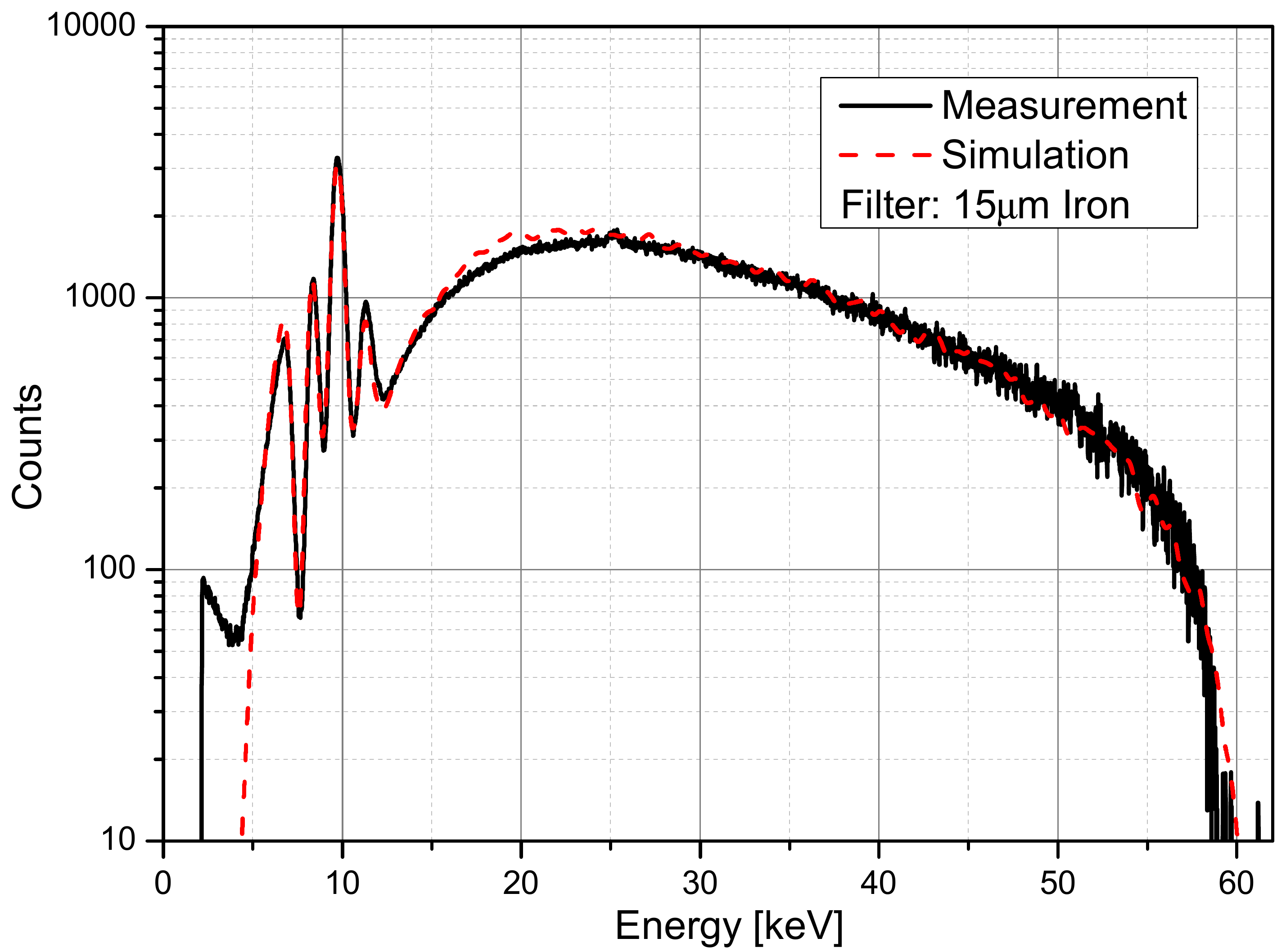}
\label{MeasSpectrum_6}
}
\caption{A comparison of the simulated and measured spectra with different filters.}
\label{Spectrum_comparison}
\end{figure*}

\subsection{Comparison of Measurements and Simulations}
\label{sec:comparison}

For a realistic comparison the simulations were convoluted with a Gaussian curve with the same FWHM as the detector.
The measured energy spectra after all available filters are in excellent agreement with the simulations, as shown in Fig. \ref{Spectrum_comparison}. Note the reasonable agreement between simulation and measurements without filter (left top panel),
which is the justification for the thin tungsten foil in Fig. \ref{Simulation_2}  representing the absorption in the collimator.
Only the low energy part differs slightly, which is expected, since 
in the simulation photons are registered, not detected by the detector. Therefore, the simulation does not include additional absorption effects in the detector,  e.g. by Compton scattering.
The small enhancement at 25.2 keV, which appears in every measured spectrum, is the K$_\alpha$ line of tin~\cite{TransEnergies}. Characteristic X-ray photons from the solder used to connect the peltier cooling in the detector capsule are responsible for this peak.

\section{Dose rate measurements with RadFETs}
To make sure that the results are reliable we crosschecked the simulation with special ionizing radiation detectors, namely RadFETs, which consist of calibrated MOSFETs, where the shift in the threshold voltage by the positive charge build-up in the silicon oxide layer has been  calibrated~\cite{RADFET1,RADFET2}. 
Table~\ref{tab:comparisonOfFilters} compares the results from the simulation  with the measurements. One observes that  filtering out the soft part of the spectrum (corresponding to the lower dose rates) improves the agreement between  measurements and simulations, as expected since the low energy part of the spectrum in less well simulated, as discussed before.
\begin{table}
  \centering
    \begin{tabular}{|c|c|c|}
    \hline
    Filter & Dose rate & Dose rate \\
          & simulation & RadFETs \\
          \hline 
           & Gy/s  & Gy/s \\
          \hline
          \hline

    Zr     & 0.24 & 0.22 \\
    \hline
    V      & 1.84 & 1.93 \\
    \hline
    Mn     & 0.86 & 0.90 \\
    \hline
    Fe     & 1.32 & 1.19 \\
    \hline
    Ni     & 1.44 & 0.99 \\
    \hline
    Air    & 6.29 & 5.95 \\
    \hline
    \end{tabular}
  \caption{A comparison of the simulated and measured dose rates after several filters in RadFets. The values correspond to a voltage of $60$ kV, $33$ mA tube current, a distance between tube and the object of 155 mm and a $SiO_2$ thickness of 130 nm.}
  \label{tab:comparisonOfFilters}
\end{table}

\section{Summary}
This paper shows a convenient method to obtain the applied dose from  X-ray irradiation in the various layers of an electronic device, namely by one calibration measurement of the total flux and a simulation of the X-ray spectra, from which  the dose rates in materials with different absorption coefficients can be calculated. The method has been crosschecked with calibrated RadFets. The difficult part is the reliable  simulation of the X-ray spectrum after the usual filtering. We found that the Geant4 X-ray simulation with the \textit{Penelope} physics package is in excellent agreement with precise measurements of the spectra using a silicon drift detector.

\section{Acknowledgments}

\begin{itemize}
\item The authors would like to thank the Institute for Nuclear Physics(IK) of the Karlsruhe Institute of Technology for lending us the silicon drift detector.
\item We thank the KEK Institute in Japan for the support with the RadFET measurements.
\item We also are grateful to the German Federal Ministry of Education and Research (BMBF) for  financial support.
\end{itemize}

\providecommand{\href}[2]{#2}\begingroup\raggedright\endgroup

\end{document}